\documentclass{aa}  

\usepackage{natbib}
\bibpunct{(}{)}{;}{a}{}{,}

\usepackage{array, multirow, graphicx}
\usepackage[dvipsnames]{xcolor}
\usepackage{wasysym}[integrals]
\usepackage{amssymb}
\usepackage{pifont}
\usepackage{siunitx}
\usepackage{booktabs, tabularx}
\usepackage{cellspace, makecell} 
\usepackage{mathtools}
\usepackage{txfonts}
\usepackage{arydshln}

\usepackage[]{hyperref}
\hypersetup{colorlinks,citecolor=black, linkcolor=black, urlcolor=black, breaklinks=true}

\newcounter{magicrownumbers}

\newcommand{\lf}{\left}
\newcommand{\rg}{\right}

\newcommand{\sistemai}{\lf\{\begin{aligned}}
\newcommand{\sistemaf}{\end{aligned}\rg.}

\newcommand{\RNum}[1]{\uppercase\expandafter{\romannumeral #1\relax}}

\begin{document}

   \title{The MUSE Ultra Deep Field (MUDF)}

   \subtitle{VIII. The cool gas distribution surrounding galaxies at redshifts $z\approx 0.5-2$}

   \author{Edoardo Santo
          \inst{1}\thanks{\email{e.santo@campus.unimib.it}}
          \and
          Michele Fumagalli \inst{1,4}
          \and
          Seok-Jun Chang \inst{2}
          \and
          Max Gronke \inst{2}
          \and
          Rajeshwari Dutta \inst{3}
          \and
          Matteo Fossati \inst{1,5}
          \and
          Mitchell Revalski \inst{6}
          \and
          Marc Rafelski \inst{6,7}
          }
          
\institute{Dipartimento di Fisica ``G. Occhialini'', Universit\`a degli Studi di Milano-Bicocca, Piazza della Scienza 3, I-20126, Milano, Italy
\and Max-Planck-Institut für Astrophysik, Karl-Schwarzschild-Stra$\rm\beta$e 1, 85748 Garching b. München, Germany
\and IUCAA, Postbag 4, Ganeshkind, Pune 411007, India
\and INAF – Osservatorio Astronomico di Trieste, Via G. B. Tiepolo 11, I-34143 Trieste, Italy
\and INAF – Osservatorio Astronomico di Brera, Via Brera 28, I-21021 Milano, Italy
\and Space Telescope Science Institute, 3700 San Martin Drive, Baltimore, MD 21218, USA
\and Department of Physics and Astronomy, Johns Hopkins University, Baltimore, MD 21218, USA}
   
\date{\today}

\abstract{We use deep MUSE data from the MUDF survey to investigate the cool gas around galaxies at redshifts $0.5 \lesssim z \lesssim 2$. We constructed two samples: one sample for a down-the-barrel analysis, probing outflows via \(\ion{Mg}{II}\) absorption against galaxy continua, and the other sample for projected galaxy pairs to examine the gas around the foreground galaxies in the transverse direction. From down-the-barrel stacked spectra, we detected blueshifted \(\ion{Mg}{II}\) absorption, indicative of outflows, in which the absorption strength increases with stellar mass and star formation rate. Lower-mass galaxies exhibit weaker absorption, but higher outflow velocities, whereas higher-mass systems retain more cool gas with slower outflows. In the transverse direction, the absorption of \(\ion{Mg}{II}\) decreases with the impact parameter, following a shallow profile. Comparing observations with radiative transfer models, we found that extrapolating an expanding halo model constrained with down-the-barrel measurements to halo scales overestimates the observed equivalent widths, likely due to the outflow geometry and the absence of the interstellar medium in the model. Our results highlight that mass, outflow geometry, and gas retention shape the cool circumgalactic medium, and that the combination of absorption and emission diagnostics provides powerful constraints on the properties of the cold halo gas.}

\maketitle

\section{Introduction}

The circumgalactic medium (CGM), that is, the diffuse, multiphase gas surrounding galaxies, is a critical component of galaxy evolution that serves as the fuel for ongoing star formation and as the receptacle for material expelled by galactic winds (\citealt{Tumlinson_2017}; \citealt{fauchergiguere2023keyphysicalprocessescircumgalactic}). In particular, cool ($T \sim 10^4\,$K) gas traced by low‐ionization transitions such as \ion{Mg}{II} offers unique insights into the baryon cycle at intermediate redshifts, since it is sensitive to inflows of recycled material and outflows of metal‐enriched winds. 
Therefore, \ion{Mg}{II} spectroscopy adds original insights that help us to understand better how galaxies evolve over cosmic time. 

Over the past decade, two complementary observational strategies have emerged for absorption studies. First, large quasar–galaxy (or galaxy–galaxy) pair surveys based on observations obtained with the Multi Unit Spectroscopic Explorer (MUSE), such as the MUSE Analysis of Gas around Galaxies (MAGG) survey \citep{10.1093/mnras/staa3147}, and the MUSE Gas Flow and Wind (MEGAFLOW) survey \citep{Schroetter_2016}, as well as catalog-based studies such as the \ion{Mg}{II} Absorber–Galaxy Catalog (MAGIICAT) \citep{Nielsen_2013_I}, and other efforts (e.g., \citealt{Chen_2010}; \citealt{Bordoloi_2011}; \citealt{Kacprzak_2012}) have exploited bright background quasars to probe \ion{Mg}{II} absorption out to hundreds of kiloparsecs around $z\sim 0.5-1.5$ galaxies. These studies have mapped the radial and azimuthal dependence of cool gas around $\sim L^\ast$ galaxies, revealing anisotropies linked to galactic outflows and accretion flows (e.g., \citealt{2019MNRAS.485.1961Z}, \citealt{Wendt_2021}). Second, “down‐the‐barrel” studies of galaxies themselves using (stacked) rest‐frame UV spectra of star‐forming galaxies to directly measure gas kinematics against the galaxy continuum detected blueshifted \ion{Mg}{II} absorption from outflows or redshifted signatures of inflows, thereby reducing the line‐of‐sight ambiguities inherent in transverse quasar probes. These analyses have revealed strong trends of an increasing absorption strength with stellar mass and star formation rate and have constrained outflow velocities and covering fractions (e.g., \citealt{Weiner_2009}; \citealt{rubin_direct_2012}; \citealt{Martin_2012}; \citealt{Rubin_2018}).
However, these large and wide-area surveys necessarily reach only moderate depths, with continuum detection limits of $m_{AB}\sim24$–25 that correspond to characteristic stellar masses $M_\star \gtrsim 10^{9.5}\,M_\odot$ at redshift $z \sim 1$. 

To directly explore the regime of lower‐mass systems, where feedback might expel gas into the halo more efficiently \citep{Dekel+1986}, deeper observations are required. In this context, integral‐field units (IFUs) have expanded the accessible parameter space by reaching down to low‐mass systems and providing a complementary, spatially resolved view of emission line structures and gas dynamics beyond the reach of traditional slit spectroscopy \citep{chang2024modelingmgiiresonance}. 
An example of the advantage of integral-field spectroscopy, which is sensitive to low-mass galaxies, is offered by the MUSE eXtremely Deep Field (MXDF). This $\gtrsim$~100~hr ultradeep field has demonstrated the power of very long exposures in characterizing the kinematics and multiphase nature of the CGM around $M_\star\sim10^9\,M_\odot$ galaxies \citep{guo_bipolar_2023}. Likewise, the MUSE Ultra Deep Field (MUDF), with its $\sim$140~hr integration in a field of $1.5\times1.5$~arcmin$^2$ (approaching 100~hr in the deepest region) and complementary HST and ground‐based imaging, offers exquisite sensitivity to cool CGM gas down to stellar masses as low as $M_\star\sim10^8\,M_\odot$. In this study, we specifically focus on characterizing the cool phase of the CGM through detailed \ion{Mg}{II} absorption down-the-barrel and transverse analysis in the MUDF to investigate how deep integral-field spectroscopy constrains the properties of galaxy halos at low masses.

The paper is organized as follows. In Sect. \ref{sect:data-red} we describe the MUDF sample and our data reduction. Section \ref{sec:dtb} presents the results of the down-the-barrel analysis, compares them with the literature, and interprets them through radiative transfer modeling. In Sect. \ref{sec:pairs} we present and discuss the results of the transverse analysis via galaxy pairs, providing a comparison with previous results and theoretical predictions. We summarize our main findings in Sect. \ref{sec:summ}.
Throughout this paper, we adopt the standard $\Lambda \rm CDM$ cosmology with $\rm H_0 = 70 \ km \  s^{-1} \ Mpc^{-1}, \Omega_m = 0.3$, and $\Omega_\lambda = 0.7$.

\begin{figure*}
\sidecaption
\includegraphics[width=6.4cm]{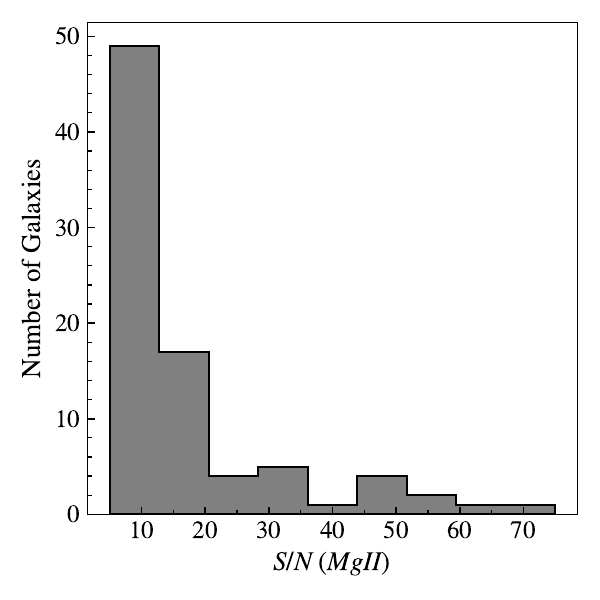}
\hfill
\includegraphics[width=6.4cm]{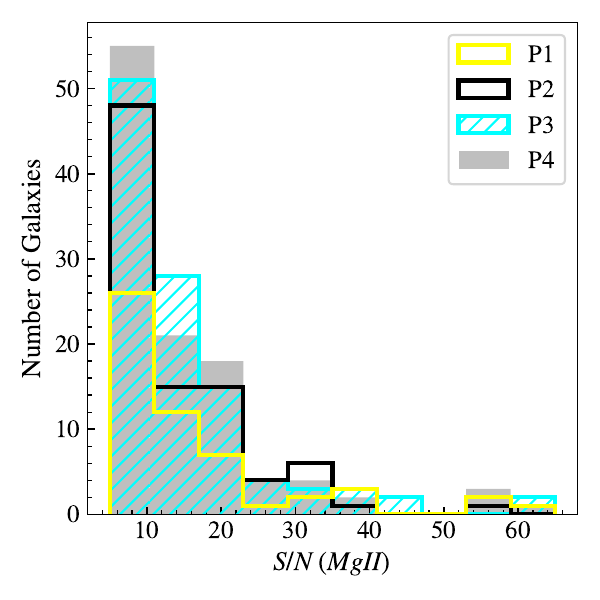}
\caption{Distribution of the continuum S/N for the MUSE spectra of the 84 galaxies included in the down-the-barrel subsample (left panel) and of the 360 galaxies used for the transverse analysis, divided into bins of impact parameter: $b \leq 40\,\rm kpc$ (P1), $40\,\rm kpc < b \leq 65\,\rm kpc$ (P2), $65\,\rm kpc < b \leq 90\,\rm kpc$ (P3), and $90\,\rm kpc < b \leq 110\,\rm kpc$ (P4; right panel).}
\label{fig:dtbsn}
\end{figure*}

\section{Observations and sample selection}\label{sect:data-red}

\begin{figure*}
        \centering
   \resizebox{\hsize}{!}{
        \includegraphics{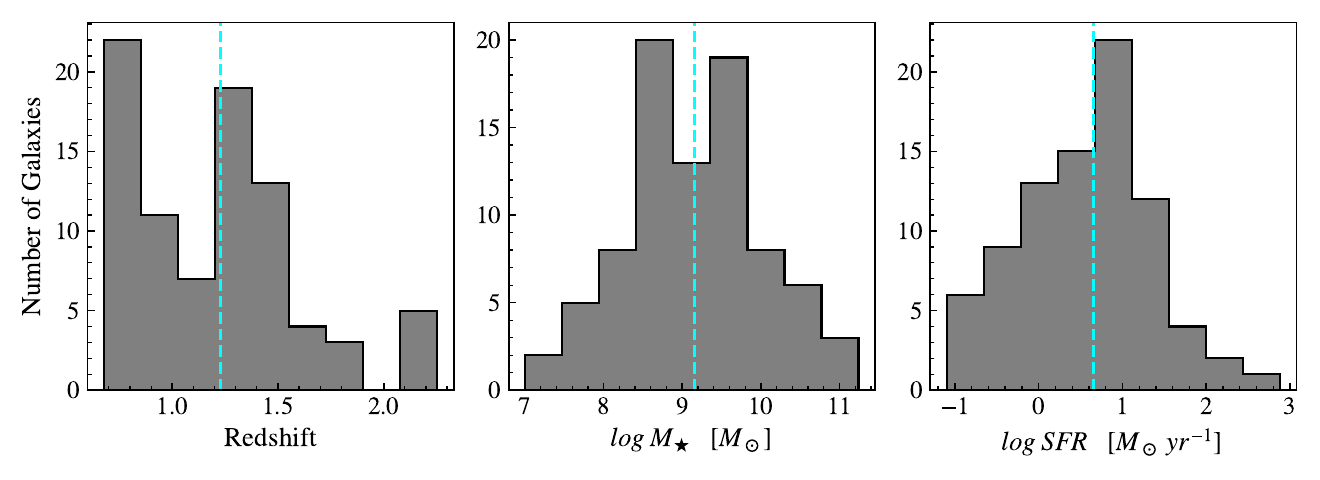}}
    \caption{Distributions of redshift (left), stellar mass (center), and SFR (right) of galaxies selected for the down-the-barrel analysis. As described in the text, these sources all have S/N > 5 in the spectral region surrounding Mg II. The dashed lines represent the median values of each distribution.}
    \label{fig:dtbphy}
\end{figure*}

\subsection{The MUSE Ultra Deep Field dataset}

Our study relies on extremely deep observations conducted with the MUSE instrument at the Very Large Telescope (VLT) \citep{bacon_muse_2010} in the $1.5\times 1.5~$arcmin$^2$ MUDF (ESO PID 1100.A-0528, PI: M. Fumagalli) with an integration time of $\sim 140$~hours in wide-field mode with extended wavelength coverage (4600-9350\AA) and spectral resolution $R\sim 2000-4000$. 
Thanks to the GALACSI adaptive optics system, the final image quality reaches a width at half-maximum of $\sim 0.73$~arcsec for point sources.
The full description of the MUSE observations and the data reduction strategy can be found in \citet{Lusso+2019} and \citet{fossati_muse_2019}. 
Hubble Space Telescope (HST) follow-up observations with 90 orbits of Wide Field Camera 3 (WFC3) F140W imaging and WFC3 G141 infrared grism spectroscopy (PID 15637, PIs: M. Rafelski \& M. Fumagalli) are also available, providing a deep galaxy catalog and a characterization of the spectral energy distributions (SEDs) from ultraviolet (UV) to near-infrared (NIR) wavelengths. The analysis of these HST observations is detailed in \citet{revalski_muse_2023}.

We relied on the catalog of continuum-detected sources provided by \citet{revalski_muse_2023}, which contains galaxies with redshifts constrained with HST and MUSE spectroscopy.
This catalog is 50 \% complete at 27.6 mag (F140W) and covers the entire area observed with MUSE, as described by \citet{revalski_muse_2023}. A quality flag represents the confidence in the redshift measurement based on the presence of high signal-to-noise (S/N) lines. We only relied on high-quality redshifts (redshift quality flag of at least 3 or above), which is assigned to 25 \% of the sources in the catalog. 
The physical properties of the galaxies, such as
stellar mass ($M_\star$) and star formation rate (SFR), were obtained by fitting the multiwavelength photometry and the MUSE spectra simultaneously with the Monte Carlo Spectro-Photometric Fitter code (see \citealp{fossati_muse_2019};  \citealp{revalski_muse_2023}; \citealp{revalski_muse_2024}). For each galaxy, we also extracted a spectrum from the MUSE cube using the HST segmentation map convolved with the MUSE point spread function and registered the wavelength to the vacuum frame. 

\subsection{Sample selection}

Our primary galaxy sample was drawn from the MUDF catalog described above and consisted of 213 sources, covering the \ion{Mg}{II} doublet within the MUSE wavelength range, with redshifts in the range $0.6<z<2.4$.
From this sample, we derived two distinct subsamples: one sample on which we performed a down-the-barrel analysis (i.e., using the galaxy continuum to probe foreground gas in absorption), and the other sample of projected galaxy pairs (i.e., where a background galaxy was used to investigate the halo gas of the foreground galaxy in the transverse direction).  

\begin{figure}
    \centering
    \resizebox{\hsize}{!}{
        \includegraphics[scale=0.8]{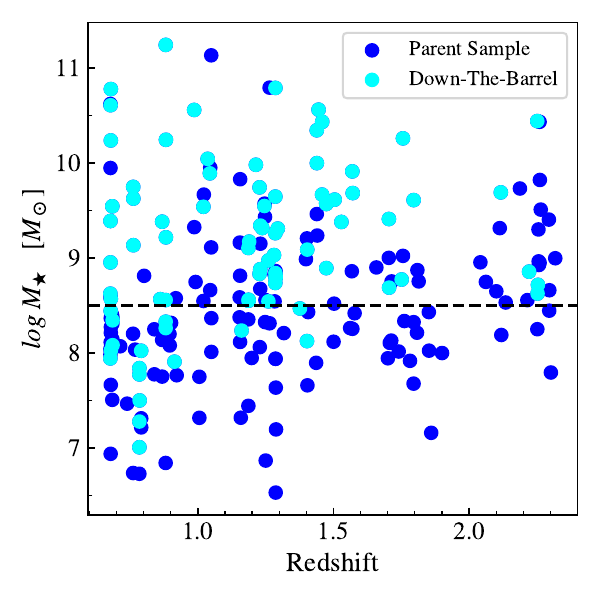}}
    \caption{$M_\star-z$ diagram of the parent sample (blue dots) and the selected subsample for the down-the-barrel analysis (cyan dots). The horizontal dashed black line at log $(M_\star/{\rm M_\odot}) = 8.5$ represents the limit we adopted for building a mass-limited sample (see Sect .~\ref{sec:dtb_1}).}
    \label{fig:zvsm}
\end{figure}

For the first set, we selected spectra with a continuum $S/N \geq 5$ per pixel, measured in a wavelength window of $\sim 25~\AA$ on either side of the expected \ion{Mg}{II} doublet (see Fig.~\ref{fig:dtbsn}). We also removed sources in which the \ion{Mg}{II} spectral window was strongly affected by skyline residuals. This subsample comprised 84 galaxies. The physical properties (mass and SFR as a function of redshift) are summarized in Fig.~\ref{fig:dtbphy}. To ensure that the subsample was representative of the parent sample, we compared the stellar mass and redshift distributions of the two samples. Fig.~\ref{fig:zvsm} shows that the down-the-barrel subsample (cyan points) follows the mass-redshift distribution of the parent sample (blue points), although it is biased toward higher stellar masses due to the selection criterion on the $S/N$. 
For the second subsample of projected galaxy pairs we used to analyze the cool gas in the transverse direction, we selected galaxies following these main criteria:
\begin{enumerate}
  \item The redshift offset between the foreground (f/g) and background (b/g) galaxies satisfies $c(z_{b/g} - z_{f/g})/(1 + z_{pair}) > 1000~\rm km~s^{-1}$, with $z_{pair} = (z_{b/g} + z_{f/g})/2$, to ensure that the background galaxy was not associated or interacting with the foreground one. This choice provided a conservative separation well above typical peculiar velocities in bound systems, excludes galaxies in the same structure, and minimizes contamination from correlated environments in absorption studies.
  \item We excluded pairs in which strong lines of the b/g galaxies, such as \ion{Fe}{II}, \ion{Al}{II}, and \ion{Al}{III}, might contaminate the \ion{Mg}{II} absorption lines of the f/g galaxies. 
  \item We only selected b/g spectra with a continuum $S/N \geq 5$ at wavelengths where we expected \ion{Mg}{II} absorption in the rest frame of the f/g galaxy. The $S/N$ distribution is shown in the right panel of Fig.~\ref{fig:dtbsn}.
\end{enumerate}

\begin{figure*}[!htbp]
        \centering
   \resizebox{\hsize}{!}{
        \includegraphics{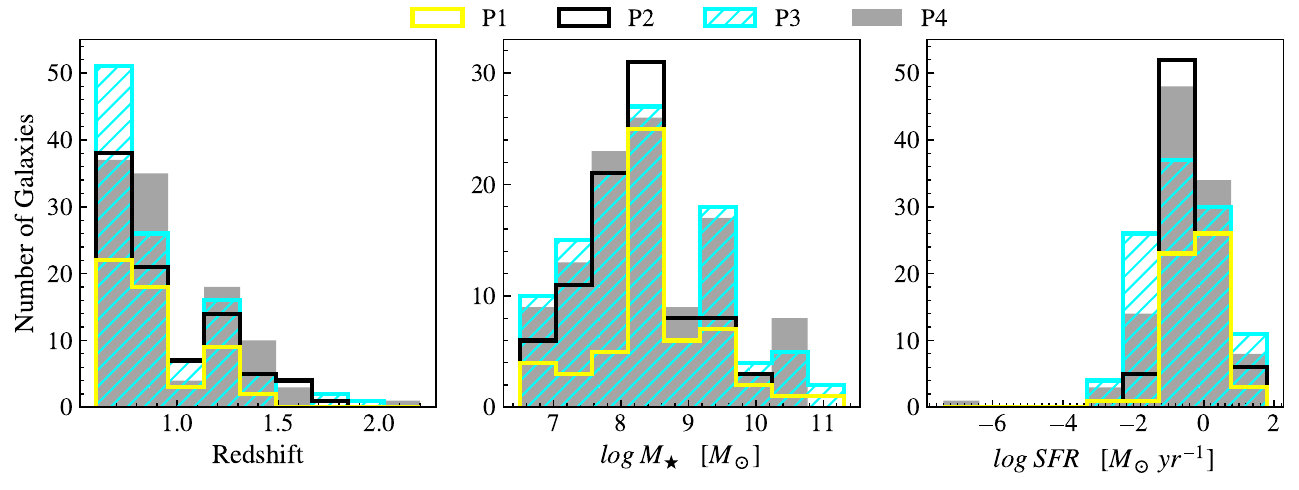}}
    \caption{Distributions of redshift (left), stellar mass (center), and SFR (right) of the f/g galaxy sample.}
    \label{fig:pair_phy}
\end{figure*}

With this selection, we identified 360 pairs, which we further split into four bins of impact parameter $b$, defined to optimize the size and S/N of measurements for each subsample: P1, with $b \leq 40~\rm  kpc$, P2, with $40~\rm kpc < b \leq 65 ~\rm kpc$, P3, with $65~\rm kpc < b \leq 90~\rm kpc$, and P4 with $90~\rm kpc < b \leq 110 ~\rm kpc$. The final pair samples included 54, 90, 108, and 108 pairs for P1, P2, P3, and P4, respectively. In Fig.~\ref{fig:pair_phy} we show the distributions of the f/g galaxy properties for each subsample. The mass distribution (middle panel) of the closest-pair sample (P1, yellow) appears to be slightly deviant with respect to the other bins.
As galaxy pairs are selected only based on chance alignment in the plane of the sky and the foreground systems share a similar redshift distribution, we do not anticipate physical reasons for a different mass distribution. We instead hypothesize that this is an effect of the small number of galaxies in this bin. To explicitly assess whether the apparent deviation is driven by limited statistics and not an intrinsic difference in the galaxy properties, we performed two-sample Kolmogorov-Smirnov (KS) tests between P1 and the other bins and between P1 and the combined sample (P2+P3+P4). In each case, the resulting p value indicated that the distributions were statistically consistent, with at most marginal differences that were not significant at conventional thresholds. In addition, we carried out a Monte Carlo exercise to further test the role of small-number statistics. We generated 1000 realizations by drawing samples of equal size to P1 from a Gaussian distribution centered at $\log(M_\star/M_\odot)=8.5$, with the dispersion matching that of the observed sample. For each realization, we performed a KS test against P1. The resulting p-value distribution shows that 95\% of realizations are statistically consistent with P1 (p value > 0.05), supporting the interpretation that the observed deviations are compatible with sampling fluctuations and do not reflect intrinsic differences.

\subsection{Continuum normalization}

To measure absorption lines, we normalized each galaxy spectrum to its continuum level, which was determined by a linear fit to the data in the spectral region on either side of the MgII doublet. In particular, we fit spectra in the windows $(2747-2767) \times (1+z)~\AA$ and $(2830-2850) \times (1+z)~\AA$, where $z$ corresponds to the redshift of the galaxy itself for the down-the-barrel analysis, while it is the redshift of the f/g galaxies for the measurements on the pairs. For the pair sample, we confirmed whether these regions in each b/g galaxy spectrum coincided with the gap caused by the \ion{Na}{} laser for AO ($5760-6010~\AA$) and/or included strong emission or absorption features of the b/g sources. When this was the case, we made minor adjustments to the continuum window to avoid incorporating extraneous features. As the spectra of galaxy pairs were analyzed through stacking, we also interpolated the spectra in the region between $-2000$ and $2000~\rm km~s^{-1}$ on a common velocity scale with a bin size of $40~\rm km~s^{-1}$, given an original rest velocity resolution between $40~\rm km~s^{-1}$ and $80~\rm km~s^{-1}$. The zero velocity was assigned to the $\lambda2796$ line of the \ion{Mg}{II} doublet.

\section{The cool gas along the line of sight to MUDF galaxies}\label{sec:dtb}

\subsection{Observed \ion{Mg}{II} stacking spectra}\label{sec:dtb_1}

First, we studied the relation between the \ion{Mg}{II} lines and the physical properties of galaxies through a down-the-barrel analysis. To do this, we split the 84 galaxies into two bins of stellar mass and two bins of SFR. To preserve a comparable number of galaxies in each bin, the thresholds were defined as the median properties of the sample at $\log (M_\star/M_\odot) \sim 9.15$ and $\log (SFR/M_\odot yr^{-1}) \sim 0.65$.

\begin{figure}
    \centering
    \resizebox{\hsize}{!}{
        \includegraphics[scale=0.8]{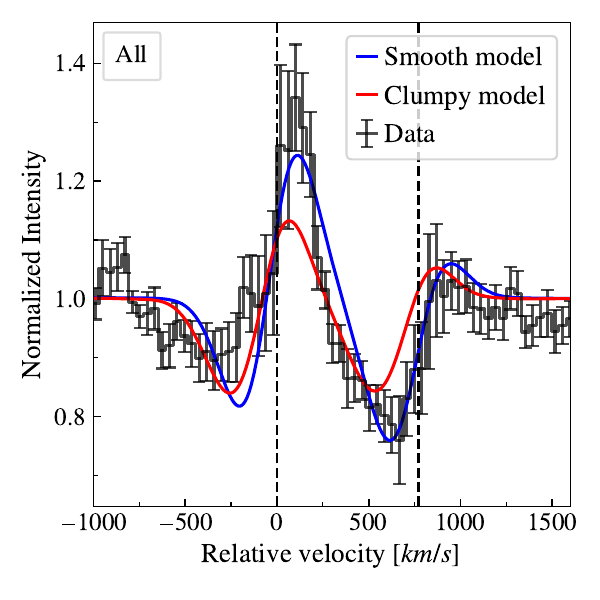}}
    \caption{Stack of all \ion{Mg}{II} down-the-barrel spectra (gray data points) and best-fit simulated spectra from \texttt{RT-scat}, assuming the gas as a smooth medium (blue line) and as a clumpy medium (red line).}
    \label{fig:all_stack}
\end{figure}

\begin{figure*}
\sidecaption
\centering
\includegraphics[width=12.9cm]{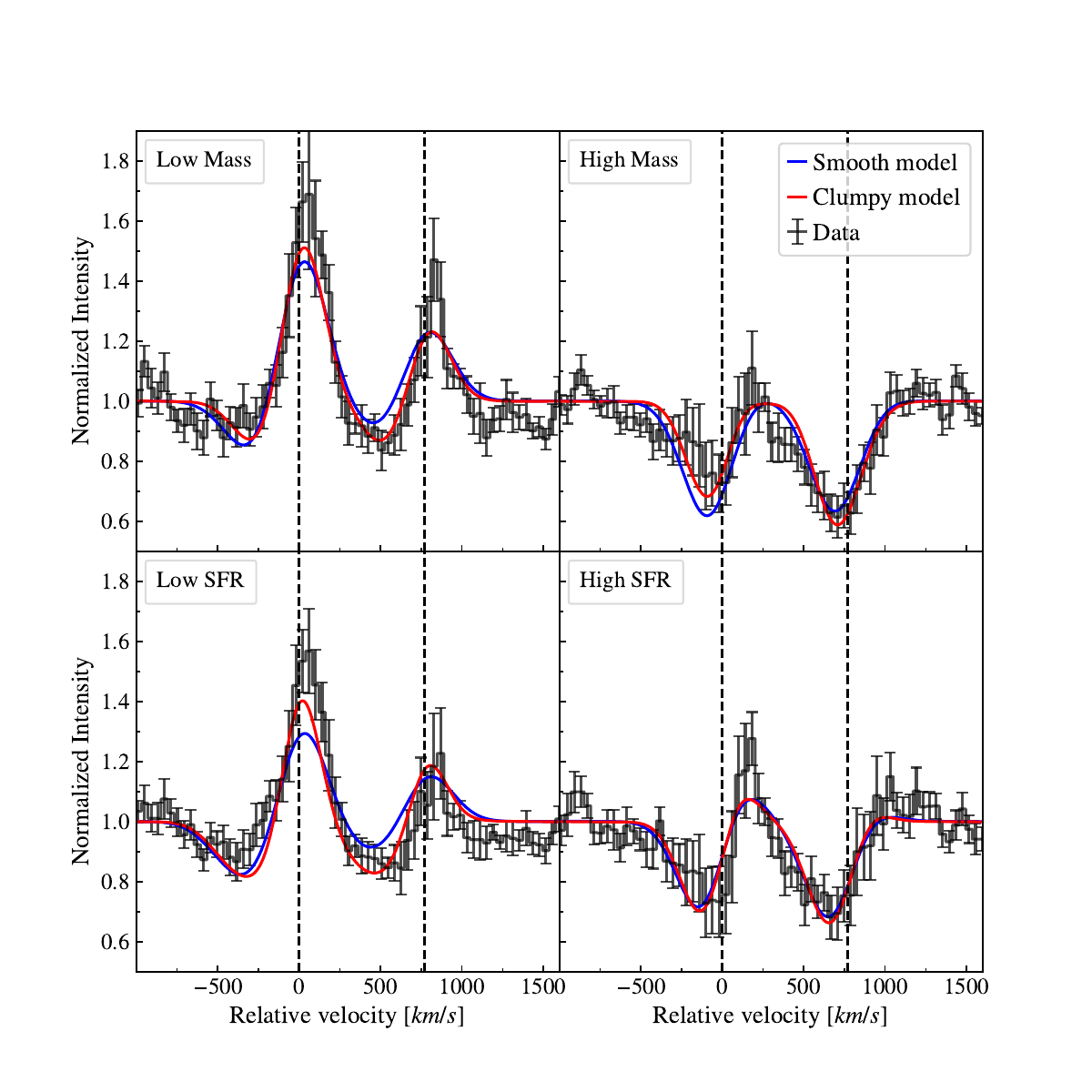}
\caption{\ion{Mg}{II} down-the-barrel stacked spectra (gray data points) and best-fit simulated spectra from \texttt{RT-scat} for the different mass (upper panels) and SFR (lower panels) bins, assuming the gas as a smooth medium (blue lines) and as a clumpy medium (red lines).}
\label{fig:stacks}
\end{figure*}

We built stacked spectra, obtained by computing the median at each wavelength, focusing on the regions around the \ion{Mg}{II} doublet
for the entire sample of 84 galaxies and for each mass and SFR subsample.
The resulting spectra are shown in Fig.~\ref{fig:all_stack} and~\ref{fig:stacks} as a function of the rest-frame velocity scale, where the zero reference is on the bluer line of the doublet. The errors were estimated as the standard deviation of the flux distributions derived by performing a bootstrap analysis with 1000 resampled spectra.
In the entire sample (Fig.~\ref{fig:all_stack}),
asymmetric blueshifted absorption in the K ($\lambda2796$) and H ($\lambda2803$) lines is readily visible. The K line peaks at a velocity of $\sim -350~\rm km~s^{-1}$ and shows a $\sim 10\%$ absorption dip from the continuum in the center of the line. The H line reaches a velocity of $\sim -100~\rm km~s^{-1}$ and is about twice as strong. An emission signature at $\sim +100~\rm km~s^{-1}$ is also evident between the absorption lines.

\begin{table}
    \caption{Best-fitting parameters (top row of each entry) and weighted standard deviations $\sigma$ (bottom row) derived through the radiative transfer model \texttt{RT-scat} applied to the down-the-barrel stacked spectra in bins of stellar mass and SFR, assuming the gas as a smooth medium.}
    \label{table:1}
    \centering
    \begin{tabular}{c c c c c c}
    \hline
     Smooth &  $\log N_{MgII} $ & $v_{exp}$ & $v_{ran}$ & $v_{emit}$ & $EW_{int}$ \\ 
      Medium &  $[{\rm cm^{-2}}]$ & $[{\rm km/s}]$ & $[{\rm km/s}]$ & $[{\rm km/s}]$ & $[\AA]$ \\
    \hline
    All & \shortstack{14.5 \\ 0.2} & \shortstack{150 \\ 39} & \shortstack{25 \\ 6} & \shortstack{25 \\ 4} & \shortstack{4.0 \\ 0.5}\\
    \hdashline
    High mass  & \shortstack{14.5 \\ 0.7} & \shortstack{100 \\ 27} & \shortstack{50 \\ 11} & \shortstack{25 \\ 59} & \shortstack{3 \\ 2}\\
    \hdashline
    Low mass & \shortstack{13.5 \\ 0.03} & \shortstack{400 \\ 41} & \shortstack{25 \\ 37} & \shortstack{25 \\ 8} & \shortstack{3.0 \\ 0.1}\\
    \hdashline
    High SFR & \shortstack{14.5 \\ 0.9} & \shortstack{150 \\ 25} & \shortstack{25 \\ 8} & \shortstack{50 \\ 33} & \shortstack{2 \\ 2}\\
    \hdashline
    Low SFR & \shortstack{13.5 \\ 0.0} & \shortstack{500 \\ 21} & \shortstack{50 \\ 43} & \shortstack{25 \\ 7} & \shortstack{2.00 \\ 0.01}\\
    \hline
    \end{tabular}
\end{table}

\begin{table}
    \caption{Same as Table~\ref{table:1}, but for a clumpy medium.}
    \label{table:2}
    \centering
    \begin{tabular}{c c c c c c c}
    \hline
     Clumpy &  $\log N_{MgII} $ & $v_{exp}$ & $v_{cl}$ & $f_c$ & $v_{emit}$ & $EW_{int}$ \\ 
      Medium &  $[{\rm cm^{-2}}]$ & $[{\rm km/s}]$ & $[{\rm km/s}]$ & & $[{\rm km/s}]$ & $[\AA]$ \\
    \hline
    All & \shortstack{14.5 \\ 0.3} & \shortstack{350 \\ 45} & \shortstack{100 \\ 17} & \shortstack{10 \\ 1} & \shortstack{100 \\ 8} & \shortstack{1.0 \\ 0.3}\\
    \hdashline
    High mass  & \shortstack{15.0 \\ 0.3} & \shortstack{50 \\ 52} & \shortstack{100.0 \\ 0.5} & \shortstack{50 \\ 17} & \shortstack{75 \\ 20} & \shortstack{6 \\ 4}\\
    \hdashline
    Low mass & \shortstack{14.5 \\ 0.3} & \shortstack{350 \\ 68} & \shortstack{100 \\ 31} & \shortstack{10 \\ 2} & \shortstack{100 \\ 12} & \shortstack{3.0 \\ 0.5}\\
    \hdashline
    High SFR & \shortstack{15.5 \\ 0.4} & \shortstack{150 \\ 58} & \shortstack{100 \\ 15} & \shortstack{20 \\ 17} & \shortstack{100 \\ 19} & \shortstack{2 \\ 2}\\
    \hdashline
    Low SFR & \shortstack{15.0 \\ 0.3} & \shortstack{500 \\ 17} & \shortstack{100 \\ 21} & \shortstack{10 \\ 1} & \shortstack{75 \\ 16} & \shortstack{2.00 \\ 0.02}\\
    \hline
    \end{tabular}
\end{table}

The high $M_\star$ and high SFR spectra in the right panels of Fig.~\ref{fig:stacks} show a deeper absorption of $30\%$, and $40\%$ at the core of the K and H lines, respectively. The line core is also more aligned with zero velocity, implying that the gas giving rise to the absorption is closer to the systemic velocity. Moreover, the emission signature disappears almost entirely in the high-mass case, but is significantly reduced (but not absent) in the high SFR subsample.    
In the low $M_\star$ and SFR cases (left panels), the emission feature instead becomes more intense, and a second weaker emission line appears at $\sim +100~\rm km~s^{-1}$ relative to the H line, giving the profiles the characteristic P-Cygni shape. To ensure that the trends observed in the stacked spectra were driven by stellar mass and not redshift, we applied a mass cut at log($M_\star/\mathrm{M_\odot}$) = 8. We then divided the subsample into low- and high-mass galaxies using the median stellar mass of this restricted sample as a threshold. The resulting stacked spectra show consistent trends with those obtained from the original division, confirming that our results are not significantly affected by a redshift dependence on the stellar mass.

Similar features have also been observed in previous studies. For example, \citet{Weiner+2009} analyzed a stacked sample of star-forming galaxies at $z \sim 1.4$ from the Deep Extragalactic Evolutionary Probe 2 (DEEP2) survey, \citet{Martin_2012} used Keck Low Resolution Imaging Spectrometer (LRIS) average spectra of galaxies at $0.4 < z < 1.4$, and \citet{2018A&A...617A..62F} worked on stacked spectra of galaxies between $0.70 < z < 2.34$ drawn from the MUSE Hubble Ultra Deep Survey. These authors reported an increasing trend of the absorption strength with stellar mass and SFR and the appearance of a P-Cygni emission line in the \ion{Mg}{II} $\lambda2796$ transition within the composite spectra of the lowest-mass galaxies in their sample.
On the more theoretical side, \citet{chang2024modelingmgiiresonance} modeled the observed \ion{Mg}{II} spectra of a comprehensive dataset of star-forming galaxies at $z \sim 1$ from the MAGG and MUDF programs, 
dividing the sample into different stellar mass bins. Consistent with our results, they found \ion{Mg}{II} emission in low-mass galaxies ($M_\star/M_\odot < 10^9$), while strong absorption becomes more prominent in higher-mass galaxies ($M_\star/M_\odot > 10^{10}$). 

\subsection{Radiative transfer modeling}

Taking advantage of available RT modeling (see also \citealt{prochaska_simple_2011}), we fit our stacked spectra using the simulated spectra generated by the 3D Monte Carlo simulation \texttt{RT-scat} (further information on the Monte Carlo implementation can be found in Sect. 2.5 of \citealt{10.1093/mnras/stae1664}). Explicitly, this model assumes two different geometries for the \ion{Mg}{II} halo, a smooth sphere and a clumpy sphere at $T = 10^4\, \rm K$, with varying expanding velocities and a point source in the center. \citet{10.1093/mnras/stae1664} assumed two types of intrinsic spectral distributions near the \ion{Mg}{II} doublet: a flat spectrum, representing the stellar continuum near $2800 \AA$, and Gaussian-like emission with a fixed flux ratio of \ion{Mg}{II} K and H lines equal to 2 characterized by a width of $v_{emit}$ and an intrinsic equivalent width $EW_{int}$. The radial outflow velocity is proportional to the distance from the central source, that is, $v(r) = v_{exp}r/R_H$, where $R_H$ denotes the outer radius. For the smooth model, the fit was performed considering five parameters: the \ion{Mg}{II} column density $N_{MgII}$, the outflow velocity $v_{exp}$, the random motion of cold gas $v_{ran}$, the intrinsic \ion{Mg}{II} emission width $v_{emit}$, and the intrinsic equivalent width of \ion{Mg}{II} emission $EW_{int}$. For the clumpy model, it also included the clumps' random velocity $v_{cl}$, and the covering factor $f_{cl}$, defined as the mean number of clumps per line of sight (for a detailed discussion of the model geometry and parameter space explored, we refer to Sect. 2 of \citet{10.1093/mnras/stae1664}).

The results of this analysis are reported in Tables~\ref{table:1} and \ref{table:2}, where we summarize the best-fit parameters obtained with a minimum $\chi^2$ for the stack of all the galaxy spectra and the stacks in each bin of mass and SFR. In Figs.~\ref{fig:all_stack} and~\ref{fig:stacks}, we also show the best fit superimposed on the stack spectra. Examining these results, we see that the \texttt{RT-scat} model is capable of capturing the main features of the observed spectra with both the medium geometries, including a varying degree of blueshift in the absorption lines as a function of stellar mass (and SFR), and the emergence of marked emission lines at lower masses. The model consistently reveals a higher column density for more massive (higher SFR) galaxies than lower masses, reflected in a more prominent absorption in the former subsample. However, we caution that the reported value is likely to be biased high by the intrinsic absorption in the interstellar medium (ISM) of the galaxies, which is included in the observed spectra, but not in the model \citep[see, e.g.,][]{rubin_direct_2012}. We also find that low–mass (low–SFR) galaxies exhibit significantly higher $v_{\rm exp}$. This suggests that their outflows are more capable of expelling gas beyond the virial radius. In contrast, in more massive systems, the outflowing material tends to stall within the halo, producing stronger absorption signatures. This was also argued by \citet{chang2024modelingmgiiresonance}.

\begin{figure}
    \centering
    \resizebox{\hsize}{!}{
        \includegraphics[scale=0.8]{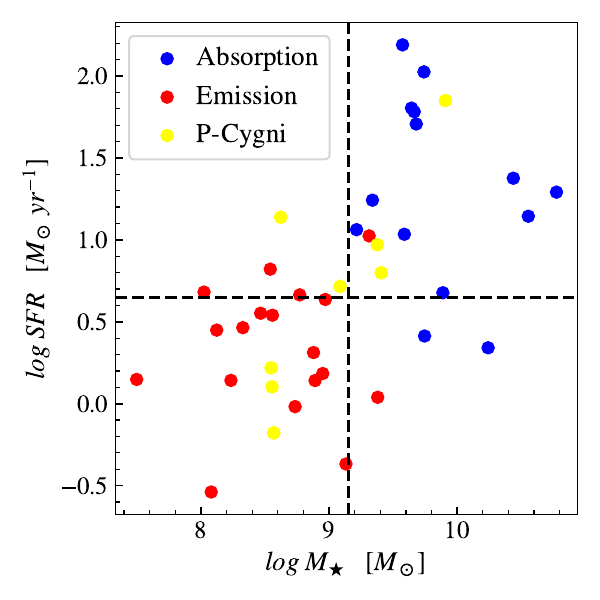}}
    \caption{Distribution of galaxies in the stellar mass and SFR plane classified as detected \ion{Mg}{II} absorption (blue), emission (red), or P-Cygni profile (yellow). The dashed black lines represent the median of the physical sample properties.}
    \label{fig:fg_properties}
\end{figure}  

\begin{figure}
    \centering
    \resizebox{\hsize}{!}{
        \includegraphics{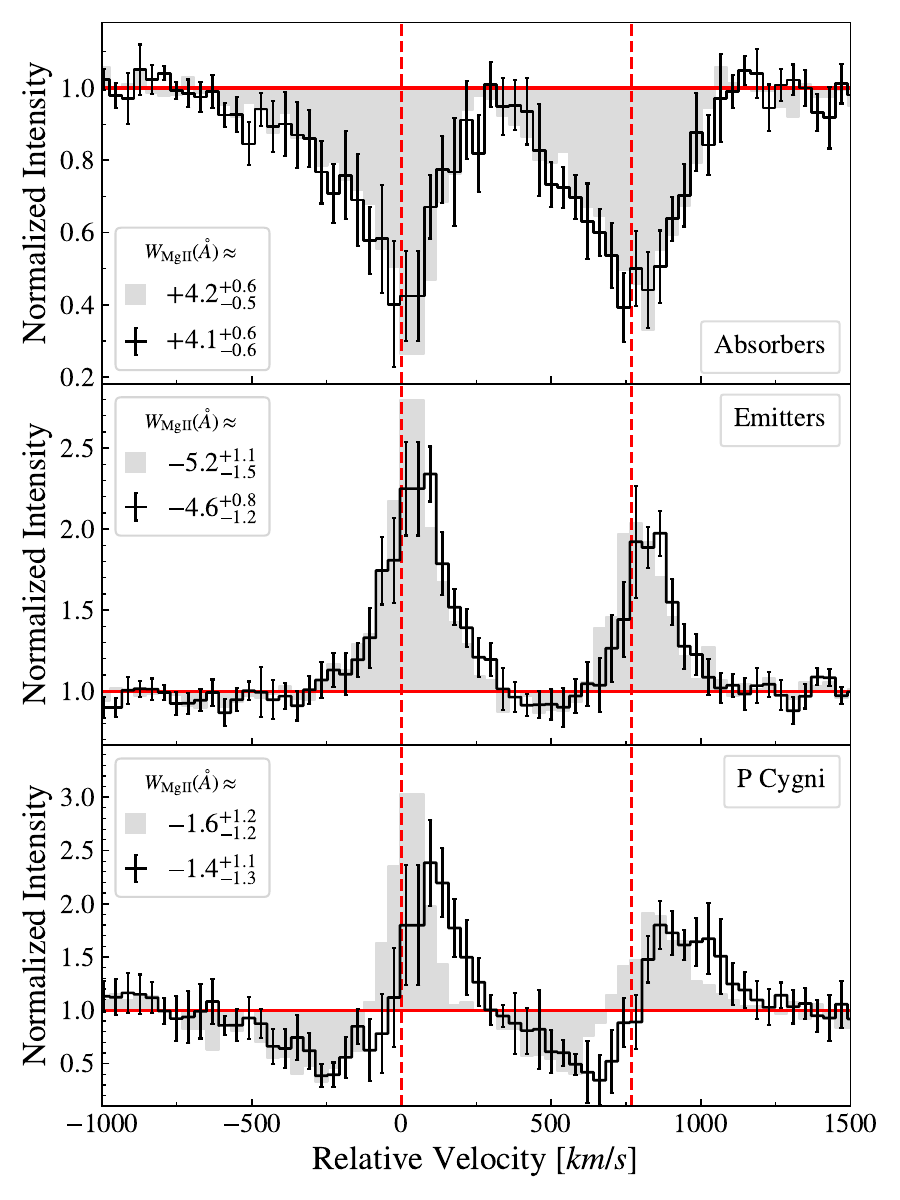}}
    \caption{Stack of spectra with detected \ion{Mg}{II} absorption (top panel), emission (middle panel), and P-Cygni (bottom panel) profiles.
    The black lines represent the stack of all the galaxies of each class relative to the systemic redshift inferred from other ISM lines. The filled gray spectra instead refer to stacks on the minimum of the K line for the absorption sample and the maximum for the emission for the emitters and the P-Cygni samples.
    }
    \label{fig:stack_foreground}
\end{figure}

\subsection{Spectral features and galaxy properties}
\label{Sec:spec_prop}

Furthermore, inspecting the individual spectra, we visually classified sources with evident \ion{Mg}{II} absorption, with emission, or displaying a P-Cygni feature, obtaining 14, 18, and 8 galaxies in each class. In Fig.~\ref{fig:fg_properties} we color-code the galaxies of each group in the stellar mass versus SFR plane. Consistent with the stacking analysis, we note that the median values of these properties neatly separate the class of \ion{Mg}{II} absorbers (with higher $M_\star$ and SFR) from the \ion{Mg}{II} emitters (characterized by lower $M_\star$ or SFR). P-Cygni profiles are more scattered throughout the stellar mass/SFR plane, but are concentrated in the middle of the distribution. Various authors have conducted similar exercises in previous works on different datasets, but reported comparable results: \citet{2013ApJ...774...50K} in star-forming galaxies at $z \sim 1$ from the DEEP2 Survey, and \citet{2012ApJ...759...26E} on a dataset obtained through a survey conducted with the LRIS-B spectrograph on the Keck I Telescope.

In Fig.~\ref{fig:stack_foreground} we further built stacked spectra for the three absorption, emission, and P-Cygni profile classes. When stacking, we resorted to two ways to determine the zero velocity of each spectrum. The first way relied on the systemic galaxy redshift as inferred from other emission lines, for example, [OII] (black lines). The second method used the \ion{Mg}{II} K line, taking the minimum for the absorption features and the maximum for the emission and P-Cygni features (filled gray lines). Within each sample, we performed a bootstrap resampling to estimate the errors of the median fluxes and to compute the distribution of the equivalent widths of the \ion{Mg}{II} doublet, $W_{MgII}$, integrated over the velocity range between $\sim -750$ and $\sim +1250~\rm km~s^{-1}$ relative to the K line. We note that partial covering of the background continuum, commonly observed in down-the-barrel Mg II absorption studies (e.g., \citealt{Rupke2005, Martin2009}), primarily affects the inference of the column densities because it alters the depth and shape of the absorption troughs. However, the integrated equivalent width is purely an observable quantity that is also well defined in the presence of nonunity covering fractions. Since our analysis exclusively relied on $W_{\rm MgII}$ as a direct measurement, the quantities we derived do not depend on any assumptions regarding the covering fraction of the absorbing gas. The implications and possible systematic effects associated with partial covering when comparing observations with models are discussed in more detail in Sect.~\ref{Sec:comp_with_models}.

We report the measured $W_{MgII}$ and error, defined as the mean and standard deviation of these distributions. For the absorption case (top panel), we note substantial agreement between the two stacking methods, reinforcing the finding discussed above: this class is mainly characterized by static or slowly outflowing gas, with no strong infalling/outflowing components.
In contrast, for the emitter case, a shift of $\sim 100~\rm km~s^{-1}$ is visible, which is indicative of a noticeable effect induced by outflows. As expected, velocity shifts contribute to a dilution effect on the line strength. However, this has little effect on the equivalent width, which remains nearly unchanged for the emission and P-Cygni features when compared to the stacked spectra relative to the ISM systemic redshift.

\begin{figure}
    \centering
    \resizebox{\hsize}{!}{
        \includegraphics{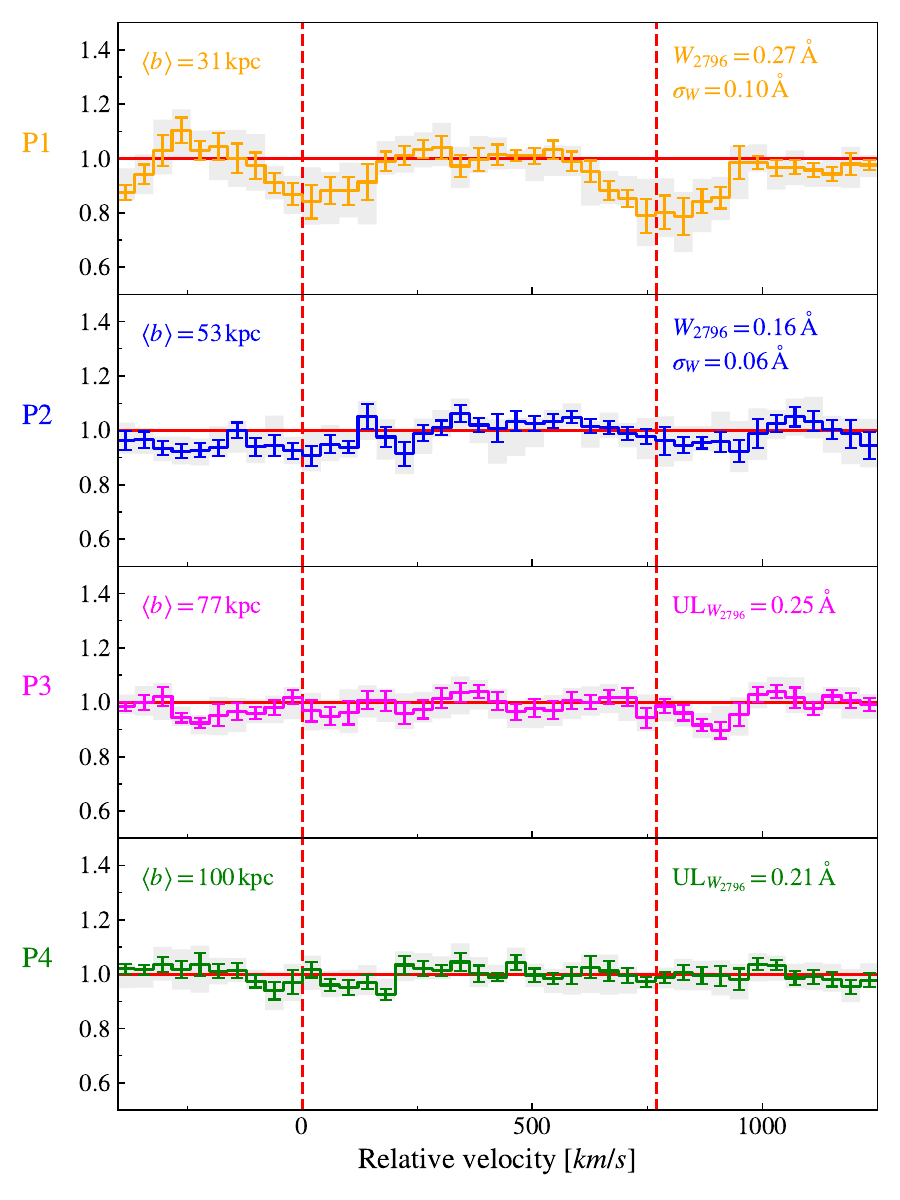}}
    \caption{Comparison of the absorption lines in composite spectra of background galaxies in pairs from P1, P2, P3, and P4 samples, from top to bottom. The standard deviation and the confidence interval for each wavelength step were estimated by making a bootstrap within each sample.}
    \label{fig:stack_background}
\end{figure}

\begin{figure*}
\sidecaption
\centering
\includegraphics[width=12.9cm]{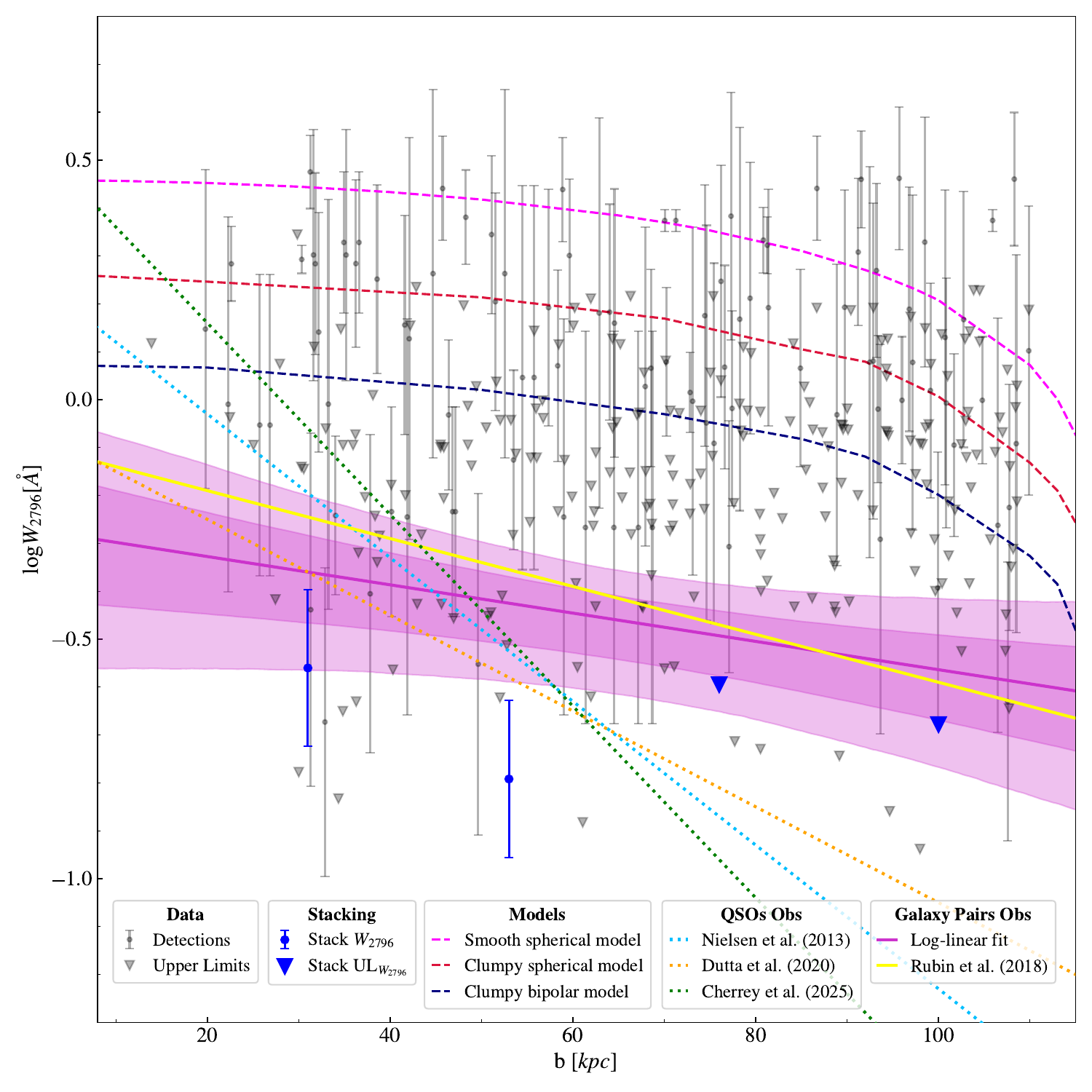}
\caption{Relation of EW ($W_{2796}$) vs. impact parameter ($b$). Measurements and upper limits ($2\sigma$) from individual pairs of f/g and b/g galaxies are plotted as black dots and triangles, respectively. The solid purple line shows the best-fitting log-linear relation. The shaded regions indicate the $1\sigma$ and $2\sigma$ bands. The red dots and triangles represent the detections and upper limits resulting from the stack in impact parameter bins. Various best-fit models from the literature are plotted \citep{Nielsen_2013, Rubin_2018, 10.1093/mnras/staa3147, Cherrey_2025}. The dashed lines represent variations in the radiative transfer model prediction for $W_{2796}$, normalized with the down-the-barrel observations and extrapolated to the halo.}
\label{fig:ew_vs_b_fit}
\end{figure*}

\section{The cool gas in the halos of MUDF galaxies via galaxy pairs}\label{sec:pairs}

\subsection{Absorption profile in galaxy halos}

After exploring the cool-gas properties of MUDF galaxies along their line of sight, we studied the cool halo gas using b/g galaxies paired with f/g ones. The depth of the MUDF and the high density of the detected galaxies that is possible with MUSE enabled this study in a small area as well. Starting from the four bins of the impact parameter defined above, we built composite spectra of b/g galaxies, shown in Fig.~\ref{fig:stack_background}. Within each sample, we performed a bootstrap resampling to estimate the errors of the median fluxes and to compute the distribution of the equivalent widths of the K line $W_{2796}$ integrated in the velocity range of $\pm~200~\rm km~s^{-1}$. We report the mean impact parameter $\langle b \rangle$, and the measured $W_{2796}$ and $\sigma_W$, defined as the mean and standard deviation of these distributions for P1 and P2. Admittedly, the detection of P2 is only marginal, but replacing the detection with a corresponding upper limit does not alter our conclusions. 
For P3 and P4 stacks, we instead report the $2\sigma$ upper limit, since the $W_{2796}$ values are consistent with zero within their uncertainties. Similarly to the down-the-barrel analysis (Sect.~\ref{sec:dtb_1}), an imposed mass cut at $\log(M_\star/M_\odot)=8$ in the transverse sample yields stacked measurements consistent within $1\sigma$ with the original values. In line with previous studies \citep[e.g.,][]{Bordoloi_2011,Nielsen_2013,Rubin_2018,10.1093/mnras/staa3147}, we observe a decreasing EW with radius at least up to $\sim 80~$kpc, at which point nondetections modulate the signal (See Fig.~\ref{fig:ew_vs_b_fit}).

\subsection{Individual detections and EW distribution}

We further complemented the study of the mean EW from the stacking analysis with the analysis of individual measurements in the spectra of b/g galaxies, as shown in Fig.~\ref{fig:ew_vs_b_fit}. The 99 individual detections scatter in the interval $W_{2796}\sim 0.3-3~$\AA\, and the relative fraction of the upper limits ($2\sigma$) increases with increasing impact parameters. However, significant detections $\gtrsim 1$\AA\ persist at these large distances, an effect that has been attributed to the increasing cross section around galaxies in groups \citep{10.1093/mnras/staa3147,dutta_metal-enriched_2021}. About 76 \% of the f/g galaxies in our sample are in groups with $\geq 3$ members, and the absorption appears to be slightly stronger in these denser environments, with a median $\log W_{2796} \sim 0.09~$\AA\ at $b > 90$~kpc, compared to $\sim -0.01~$\AA\ for isolated galaxies.

To characterize the resulting dependence of the EW on the impact parameter, we modeled our dataset assuming a log-linear dependence of $W_{2796}$ on $b$ as in \citet{Nielsen_2013}, for instance,
\begin{equation}
    log \frac{W_{2796}}{\AA} = q + m \times \frac{b}{kpc}.
\end{equation}
Following the method of \citet{Chen_2010}, the likelihood function for this model can be written as
\begin{align*}
    \mathcal{L} & \left(\overline W\right) =  \left( \prod_{i=1}^n 
    \frac{1}{\sqrt{2\pi s_i^2}} 
    \exp \left\{ -\frac{1}{2} \left[\frac{\left( W_i - \overline W \right)}{s_i} \right]^2\right\} \right) \nonumber \\
    & \times \left( \prod_{i=1}^m \int_{-\infty}^{W_i} 
    \frac{dW'}{\sqrt{2\pi s_i^2}} 
    \exp \left\{ -\frac{1}{2} \left[\frac{\left( W' - \overline W \right)}{s_i} \right]^2\right\} \right) ,
\end{align*}
where $W_i$ represents the $\log W_{2796}$ value for each measurement $i$, and $\overline W$ equals the value of $\log W_{2796}$ given by the model at each $b$. The first product includes all detected systems, and the second product includes the systems for which our constraint on $\log W_{2796}$ is an upper limit. We further assumed that the variance of the data around the mean relation is a Gaussian with two components
\begin{equation*}
    s_i^2 = \sigma_i^2 + \sigma_C^2\:,
\end{equation*}
where $\sigma_i$ is the measurement uncertainty in $W_{2796,i}$, and $\sigma_C$ is an additional factor that accounts for intrinsic scatter in the relation.

We sampled the posterior probability density function (PPDF) for this model using the Markov chain Monte Carlo (MCMC) method implemented in \texttt{emcee} \citep{Foreman+2013}. We adopted uniform probability densities for the intervals $-5.0 < m < 5.0$, $-10.0 < q < 10.0$, and $0 < \sigma_C < 10.0$ as priors. 
Markov chains generated by 100 walkers taking 6000 steps each (discarding the first 1000 steps) provide an exhaustive sampling of the PPDF in each parameter dimension \citep[see][]{Rubin_2018}.
We defined the best value of each parameter and its uncertainty interval as the maximum and the 16th and 84th \%iles of the inferred PPDFs. 

The resulting relation is flat, with a slope of $m = -0.003\pm0.002$, an intercept of $q=-0.3\pm0.1$, and an intrinsic scatter of $\sigma_C = 0.5\pm0.1$. In Fig.~\ref{fig:ew_vs_b_fit} we show the fit result (in purple) with the $1\sigma$ and $2\sigma$ bands compared with the fit previously performed on other datasets from quasar \citep[dotted lines,][]{Nielsen_2013, 10.1093/mnras/staa3147, Cherrey_2025} or galaxy b/g \citep[solid line,][]{Rubin_2018} sightlines. 
The inferred dependence of the EW on the impact parameter appears to be consistent with the result of the stacking analysis discussed above. Our trend is consistent with the trend reported by \citet{Rubin_2018}, whose results were also based on stacking background galaxy spectra, but are flatter than the others. 
As noted by \citet{Rubin_2018}, differences in slope among the datasets might arise from selection effects, such as the inclusion of isolated or group-member galaxies. 
This effect was tested by \citet{10.1093/mnras/staa3147}, who found a flattening relation when group galaxies were included. However, this difference seems marginal in their analysis. As noted by \citet{rubin_galaxies_2018}, other factors such as the interval of $b$ considered for the fit and the different nature of quasar versus galaxy spectroscopy also play a role in modulating the trend. In particular, background quasar sightlines probe narrow pencil beams (of a few parsecs, smaller than typical \ion{Mg}{II} cool clouds), while background galaxies sample regions that are several kiloparsecs across, averaging over multiple clouds. As a result, the covering fraction measured and the corresponding transverse profile from galaxy–galaxy and galaxy-quasar pairs should be compared with caution.

\subsection{Comparison with model predictions}
\label{Sec:comp_with_models}

Next, we compared the prediction of the halo model described above, constrained down the barrel, with the observations of the EW distribution measured in the transverse direction. To do this, we reconstructed the \ion{Mg}{II} column density and outflow velocity with the ``all'' parameters from Table~\ref{table:1} (smooth model) and from Table~\ref{table:2} (clumpy model), setting the remaining parameters to the values fixed by \citet{chang2024modelingmgiiresonance}. $R_{H}$ was set to $120~\rm kpc$ to include the maximum impact parameter we observed. The predicted spectra were constructed as follows. 

For the smooth model, we computed the intersection of the line of sight with the spherical volume at each selected impact parameter. This path was then divided into segments of equal length. For each segment, we calculated the line-of-sight velocity and assigned an optical depth along with a Gaussian velocity profile centered on that velocity. The width of each Gaussian was set by the thermal broadening of \ion{Mg}{II} at $10^4$~K. The complete absorption spectrum was obtained by summing the contributions from all segments along the line of sight.

For the clumpy model, the approach was similar, with a few additions. We estimated the column density of each clump as $N_{MgII,cl} = \frac{3 \ N_{MgII}}{4 \ f_c}$ and determined the number of clumps intersected at each impact parameter as $2 \ f_c \ \sqrt{1 - (\frac{b}{R_{H}})^2}$. For each clump, we calculated the line-of-sight velocity. Different from the smooth case, we added a random motion term to the radial outflow velocity, modeled as a Gaussian deviate $\mathcal{N}$(0,1) scaled by the clump random motion $\nu_{cl}$ obtained from the fit. Then, we assigned an optical depth and a Gaussian velocity profile to each clump, following the same method as in the smooth case.  

The dashed lines in Fig.~\ref{fig:ew_vs_b_fit} represent the $W_{2796}$ versus $b$ relation derived from these models. In general, the halo model predicts an almost flat profile, consistent with the fit result, followed by a drop at the assumed outer edge of the halo. However, this cutoff at the external radius of the halo is an inherent feature of the model. In contrast, the CGM is not expected to have sharp edges, and the contribution of additional halos becomes progressively more relevant with increasing radius.

The smooth model lies in the upper envelope of the detected systems, but exceeds the best fit of the observed EW. The clumpy model, which describes the expected \ion{Mg}{II} distribution in the halo better \citep[e.g.,][]{schroetter_muse_2019}, predicts a lower value of EW, but lies consistently above the fit. This excess likely arises from simplifying assumptions in the model.

First, the halo models omit an ISM component, which is present in the data. Second, they assume either a uniform gas density or a constant number of clumps, whereas in reality, the clumpiness and the overall density of cold gas both decline with galactocentric radius. Beyond the missing ISM contribution and radial density gradient, the biconical nature of the outflows differs from the spherical one assumed by the model and also depends on the inclination of the line of sight with the galactic plane. Previous observations have indeed revealed an anisotropic distribution of circumgalactic cold gas (e.g., \citealt{schroetter_muse_2019}; \citealt{2013Sci...341...50B}; \citealt{Kacprzak_2012}; \citealt{10.1111/j.1365-2966.2012.21114.x}; \citealt{Ho_2019}), which likely amplifies the discrepancy between spherical‐model predictions and sightline measurements. As a result of these combined effects, we expect a broad distribution in column density with a lower mean than measured in the down-the-barrel line of sight, which more readily crosses the outflow component. Similar geometric considerations are also expected to affect the effective velocity along the line of sight. 

Following this idea, using the same parameters as obtained with the clumpy spherical model, we considered a bipolar wind geometry with an opening angle of $30$~deg \citep{chang2024modelingmgiiresonance,guo_bipolar_2023}. By generating random lines of sight for each impact parameter, we found a covering fraction of $\sim 14\%$, which we used as a scaling factor for the column density. The result, shown as the dashed blue line in Fig.~\ref{fig:ew_vs_b_fit}, aligns better with the observations.

A further caveat concerns the role of the covering fraction in shaping the predicted absorption strength, as introduced in Sect.~\ref{Sec:spec_prop}. Since our radiative-transfer framework assumes a homogeneous statistical distribution of clumps for the clumpy model by construction, the effective area-covering fraction is encoded in the mean number of intercepts per sightline and is therefore fixed at all radii. The question arises whether the derived equivalent width distributions depend on the form of the background beams (either extended, as for galaxies, or compact, like in quasars). To assess this effect, we carried out a simple numerical experiment in which, instead of using single skewers, we simulated beam-like sightlines sampling a finite transverse area. We then compared the resulting equivalent widths with those obtained from infinitesimal skewers.

For the smooth model, the two approaches yield identical results. This is expected, as averaging over multiple rays in a medium with uniform density is equivalent to evaluating a single line of sight through that medium. For the clumpy model, where the number of clumps per sightline is drawn from the mean parameter $f_c$, beams produce EW values that are only marginally lower than those from skewers. This small difference arises from the averaging over the finite beam area, but remains negligible because the clump population is defined through a mean statistical covering that already captures the effective probability of intercepting gas. In this modeling framework, the beam-covering fraction is therefore inherently contained in the definition of $f_c$, and its impact on the mean predicted absorption is minimal.

This test confirms that within the assumptions of the present RT setup, adopting beams rather than skewers does not significantly modify the predicted equivalent-width profiles. However, it also highlights that the model has simplistic characteristics, that is, it does not capture any radial variation in the covering fraction, which is expected observationally. A more sophisticated treatment of the spatial distribution of cool gas in the models will be the subject of future work.

\section{Summary and conclusions}\label{sec:summ}

Our study leveraged deep MUSE observations from the MUDF survey to investigate the cool-gas properties in and around galaxies at $0.5 \lesssim z \lesssim 2$. Using a catalog of continuum-detected sources with high-quality redshifts from \citet{revalski_muse_2023}, we constructed two samples of galaxies with masses between $10^7$ and $10^{11.5} M_\odot$ and an SFR between $10^{-7}$ and $10^3~\rm M_\odot~yr^{-1}$: one sample for a down-the-barrel analysis to probe foreground gas in absorption through the galaxy continuum, and the other sample for projected galaxy pairs to investigate halo gas in the transverse direction.

For the down-the-barrel analysis, we selected 84 galaxies with $S/N \geq 5$ and minimum skyline contamination to examine their \ion{Mg}{II} absorption features. In the stacked spectra of two mass and SFR bins, we detected asymmetric blueshifted absorption features \ion{Mg}{II} indicative of outflows. The absorption strength increases with stellar mass and SFR, whereas P-Cygni profiles are more prevalent in lower-mass, lower-SFR galaxies. We modeled the spectra with the 3D Monte Carlo radiative transfer simulation \texttt{RT-scat} \citep{10.1093/mnras/stae1664}, finding that high-mass galaxies exhibit higher column densities and lower outflow velocities. In contrast, low-mass galaxies show weaker absorption, but faster outflows, suggesting more efficient gas expulsion in low-mass systems \citep{chang2024modelingmgiiresonance}.

For the projected pair sample, we analyzed 360 galaxy pairs, with b/g spectra $S/N \geq 5$ grouped by impact parameter and avoiding the contamination of b/g strong lines, to study the transverse distribution of \ion{Mg}{II} absorption in the CGM. The \ion{Mg}{II} absorption strength exhibits a declining radial profile
around galaxies. The dependence of $W_{2796}$ (including $2\sigma$ upper limits for nondetections) on the impact parameter is flatter than reported in previous studies using quasar spectroscopy \citep[e.g.,][]{Nielsen_2013,dutta_metal-enriched_2021}, but agrees with the analysis of galaxy spectroscopy by \citet{Rubin_2018}. Various effects related to the environment of galaxies, the interval of the impact parameter considered for the fit, and the different nature of quasar versus galaxy spectroscopy contribute to this discrepancy.

Next, we compared the \citet{10.1093/mnras/stae1664} halo model predictions for a smooth and a clumpy medium, constrained through a down-the-barrel measurement, with observations of the EW distribution in the transverse direction. Using the best estimates of the column density and outflow velocity for the entire sample, we reconstructed the halo profile. The model predicts an almost flat EW in line with the data, but the predicted EW exceeds the observed values, especially for the smooth model. While the clumpy model agrees better, it still overshoots the observations considerably. We interpret this discrepancy as caused by the lack of an ISM component in the model and differences in the outflow geometry and galactic plane inclination. Indeed, we showed that the resulting equivalent width shifts in the direction of the observations when we included the geometric dilution induced by a biconical outflow with an opening angle of 30~deg. 

Our analysis offers new insight into the distribution of cool gas in the CGM of galaxies, considering the absorption properties along the line of sight to their star-forming regions and in the transverse direction inside their halos, using a deep dataset that reaches masses as low as $M_\star\lesssim 10^8~\rm M_\odot$. 
Thus, this study complements the work of \citet{chang2024modelingmgiiresonance}, who employed a similar dataset to combine absorption diagnostics along the line of sight with emission measurements in the halos. We can thus directly compare the findings obtained by relying on two different probes inside the halo.

Applying radiative transfer modeling to the down-the-barrel absorption/emission spectrum and the \ion{Mg}{II} surface brightness in emission inside halos, \citet{chang2024modelingmgiiresonance}
constrained the key halo parameters in Tables~\ref{table:1} and ~\ref{table:2}. They observed that low-mass galaxies display prominent \ion{Mg}{II} emission and high outflow velocities in the core ($b < 10 ~\rm kpc$) and nearest the halo regions ($10 < b < 30 ~\rm kpc$), and they concluded that these features can be explained by fast-moving outflowing gas components. Conversely, high-mass galaxies tend to show strong core absorption, with lower outflow velocities indicative of abundant, relatively static cold gas-accompanied by more extended halo emission. 

Similar conclusions are confirmed by our down-the-barrel analysis alone: low-mass galaxies are better modeled by a lower column density and a higher velocity than more massive systems. This agrees well with the picture that more massive galaxies retain more cool gas in their halos, whereas lower-mass objects can more effectively push gas outward under the effect of winds. It might be concluded that down-the-barrel information alone can constrain the system fully. However, we showed that extrapolating the information recovered along the line of sight to galaxies to their halos yields a higher EW than observed. 
This discrepancy can be mitigated by the effects of an asymmetric wind geometry, as often invoked in \ion{Mg}{II} absorption studies \citep{schroetter_muse_2019}. 
The transverse component in absorption is thus still critical to add information about the spatial distribution and geometry of the cool phase of the outflow \citep[see also][for a complementary approach in emission]{guo_bipolar_2023}. 

Using complementary techniques in deep datasets, we conclude that the mass is a fundamental parameter in shaping the properties of cool gas inside halos, but geometry still plays a critical role. As various diagnostics in absorption and emission become more available for the same samples, the constraining power of refined radiative transfer models will yield a comprehensive picture of the cold CGM.

\begin{acknowledgements}
We thank Kate H.R. Rubin, Rongmon Bordoloi, and Davide Gerosa for discussions that improved this work.
This work is based on observations collected under ESO programme ID 1100.A-0528. Based on observations with the NASA/ESA Hubble Space Telescope obtained from the MAST Data Archive at the Space Telescope Science Institute, which is operated by the Association of Universities for Research in Astronomy, Incorporated, under NASA contract NAS5-26555. Support for program numbers 15637 and 15968 was provided through a grant from the STScI under NASA contract NAS5-26555. These observations are associated with program numbers \href{https://archive.stsci.edu/proposal_search.php?mission=hst&id=6631}{6631}, \href{https://archive.stsci.edu/proposal_search.php?mission=hst&id=15637}{15637}, and \href{https://archive.stsci.edu/proposal_search.php?mission=hst&id=15968}{15968}.
\end{acknowledgements}

\bibliographystyle{aa}
\bibliography{MyBib_over}
\end{document}